\documentclass[%
reprint,
 amsmath,amssymb,
 aps,
pra,
floatfix,
]{revtex4-2}
\usepackage{graphicx}
\usepackage{dcolumn}
\usepackage{bm}
\usepackage{subfigure}%
\usepackage{dcolumn}
\usepackage{bm}
\usepackage[mathlines]{lineno}
\usepackage[utf8x]{inputenc} 
\usepackage[T1]{fontenc}
\usepackage{amsmath}
\usepackage{braket}
\usepackage{ucs}

\begin{document}

\preprint{PRA}

\title{Efficient quantum state preparation using Stern-Gerlach effect on cold atoms \\}

\author{Vivek Singh}
\email{viveksingh@rrcat.gov.in}
\author{V. B. Tiwari}%
\author{S. R. Mishra}
\affiliation{Laser Physics Applications Section, Raja Ramanna Centre for Advanced Technology, Indore-452013, India\\
 Homi Bhabha National Institute, Anushaktinagar, Mumbai-400094, India 
}%




\date{\today}

\begin{abstract}
 The Zeeman hyperfine state dependent force in a Stern-Gerlach (SG) experiment has been exploited to separate and detect atoms having different Zeeman hyperfine states in a cold atom cloud. Utilizing this SG technique, we have made the quantitative estimate of atoms in different Zeeman hyperfine states in an atom cloud, which has been helpful in optimizing the optical pumping of atoms for efficient  preparation of atomic state. Employing an optimized optical pumping, nearly $92 \% $ of cold $^{87}Rb$ atoms from a grey magneto-optical trap (G-MOT) on atom-chip have been optically pumped to the trappable Zeeman hyperfine state $\ket{F=2, \, m_{F} = +2}$. These optically pumped atoms have been trapped in an Ioffe-Pritchard magnetic trap near the atom-chip surface. 

\end{abstract}

\maketitle

\section{Introduction}
An atom-chip \cite{folman, folman1, den,fort, reichel, eckel, Reichel1} with micron sized wire structure patterns on a reflecting surface provides platform for trapping and manipulation of atoms on miniaturized scale. The manipulation of cold atoms on a miniaturized scale using atom-chip has been demonstrated in a number of ways including atom traps \cite{reichel}, atom guides\cite{muller, Dekker}, beam splitters \cite{casset, machluf} and Bose-Einstein condensations (BECs) \cite{hansel, Farkas, Ott}. Usually, a magneto-optical trap (MOT) on atom-chip is prepared by reflecting the MOT-beams (cooling and re-pumping laser beams) on atom-chip surface in presence of a quadrupole magnetic field produced by the bias coils (outside the chamber) and an U-shaped wire (kept behind the atom-chip). This MOT is usually called "U-MOT". After initial loading of the U-MOT, the quadrupole magnetic field and cooling laser beam detuning are increased which brings the atom cloud closer to  chip surface but with reduced temperature and increased density. This MOT is called grey-MOT (G-MOT). Typically the cold $^{87}Rb$ atoms in a MOT cloud are equally distributed among different magnetic sub-states of hyperfine state F=2. For the magnetic trapping, however, the state $\ket{F=2, \, m_{F} = + 2}$ is a preferred state. Therefore, maximum number of atoms are to be prepared in this state to maximize the number of atoms in the magnetic trap \cite{mishra}. The presence of other states in the MOT atom cloud, particularly un-trappable states $\ket{F=2, \, m_{F} = -2,-1,0}$, results in a loss of the cold atoms from the magnetic trap. This loss can be as large as $\sim$ 60 $\%$ of the atoms from the MOT cloud \cite{david}. Thus, the preparation of atoms in the trappable state using optical pumping is an important step before the magnetic trapping.\\ 

The efficiency of optical pumping of atoms to the trappable state ($\ket{F=2, \, m_{F} = +2}$) can be quantified by applying Stern-Gerlach technique which can separate atoms in different Zeeman hyperfine states \cite{david}. In this work, we have used this SG technique to employ the Zeeman state dependent magnetic force (Stern-Gerlach force) on cold $^{87}Rb$ atoms in a MOT to spatially separate and detect them in different Zeeman hyperfine states. In our experiments, by varying the temperature of the cold atom cloud and duration of Stern-Gerlach magnetic field, the spatial separation of cold atoms in the different Zeeman states was optimized to accurately detect atoms in these states. This has allowed to estimate accurately the optical pumping efficiency in our experiments for pumping atoms to a trappable Zeeman hyperfine state ($\ket{F=2, \, m_{F} = +2}$). After optimizing the power and pulse duration of optical pumping pulse, $\sim 92 \% $ of atoms from a G-MOT on the atom-chip were transferred to trappable state $\ket{F=2, \, m_{F} = +2}$. The optically pumped atoms were trapped in external Ioffe-Pritchard (IP) magnetic trap near atom-chip surface formed by magnetic fields due to a current carrying Z-shaped copper wire and bias coils.\\

\section{Experimental setup}
\begin{figure}[ht]
\centering
\includegraphics[width=9.0 cm]{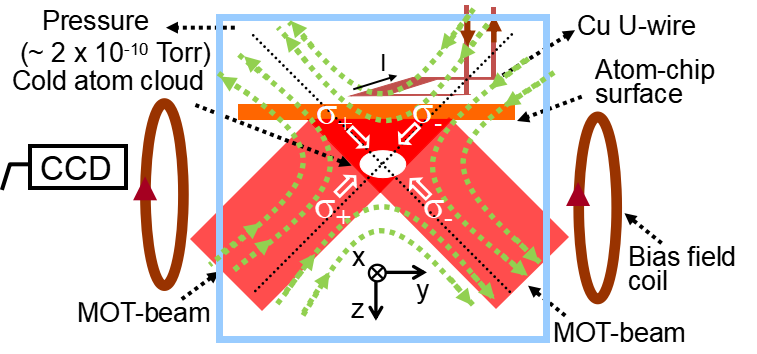}
\caption{ Schematic of the atom-chip setup for MOT formation, with two beams in reflection geometry forming four MOT beams in yz-plane, two counter propagating MOT beams in $\pm$ x-axis (not shown in figure), a U-shaped copper wire and bias coils (y-direction bias coils shown and z-direction coils are not shown). The dotted curves show the magnetic field lines. The MOT is formed $\sim$ 6 mm below the atom-chip surface. }
\end{figure}
The schematic of the experimental setup used in this work is shown in Fig. 1. The details of the experimental setup can be found in our earlier published works \cite{vivek1, vivek2}. The experimental setup consists of an octagonal vacuum chamber which is connected to vacuum pumps through a six way cross. A turbo molecular pump (TMP) (77 l/s), a sputter ion pump (SIP) (300 l/s) and a titanium sublimation pump (TSP) are the part of vacuum system used for creating an ultrahigh vacuum (UHV) with a base pressure of $\sim$ 2.0 $\times 10^{-10}$ Torr in the chamber. The Rb-dispensers are fixed on the UHV side rods of a feed-through and this feed-through is inserted in the chamber such that dispensers are placed at a distance of $\sim$ 17 cm from the center of the octagonal chamber. The Rb vapor is generated in the chamber when a current of $\sim$ 3 A is passed through the feed-through rods. The Rb-vapor generation  results in increase in the pressure in the UHV chamber to $\sim$ 3.5 $\times 10^{-10}$ Torr.\\ 

The atom-chip is mounted on a special mount and feed-through system. This is inserted inside the UHV chamber such that atom-chip active surface is facing vertically downwards (along z-axis in Fig.1). Four independent MOT beams (cooling laser beams mixed with re-pumper laser beams) are used to form U-MOT with quadrupole field produced by U-shaped external copper wire (cross-section: 1 mm $\times$ 10 mm ) and bias coils, as shown in Fig. 1. On the back surface of atom-chip, on the top of copper U-wire, a Z-shaped copper wire (cross section : 1 mm $\times$ 1 mm) is kept to produce Stern-Gerlach magnetic field. This wire is also used for magnetic trapping of atoms in Ioffe-Pritchard (IP) magnetic trap. The field due to this Z-wire and field due to the bias coils provide the net field required for an external Ioffe-Pritchard (IP) magnetic trap near the chip surface. The 2f- fluorescence imaging technique \cite{serre} has been used to image the cold atom cloud in x-z plane using a digital CCD camera.\\

\begin{figure}
\centering
\includegraphics[width=9.0 cm]{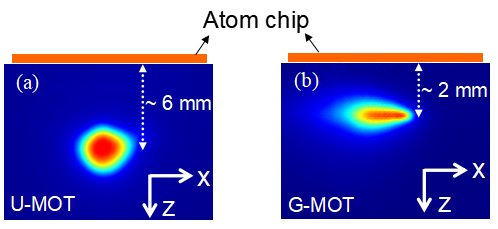}
\caption{ CCD images of cold $^{87}Rb$ atoms in (a) U-Magneto-optical trap (U-MOT) and (b) Grey-magneto optical trap (G-MOT). }
\end{figure}

A FPGA based electronic controller is used to generate digital as well as analog pulses to control various equipment like power supplies, acousto-optic modulators, CCD camera, mechanical shutters etc to implement various experimental events, from MOT-loading to magnetic trapping and cloud imaging, in a sequential manner. The first step in this sequence is to load U-magneto-optical trap (U-MOT) for $^{87}Rb$ atoms. The duration of this satge is $\sim$ 20 s.  A current of 60 A is flown in U-wire and bias fields of $\sim$ 10 G in y-direction and $\sim$ 1.5 G in z-direction is applied during U-MOT formation. The cooling laser frequency is kept - 14 MHz red-detuned from the cooling transition of $^{87}Rb$. The U-MOT cloud is formed $\sim$ 6 mm from the atom-chip surface, as shown in Fig. 2 (a). The number and temperature of cold atoms in U-MOT were $\sim 1 \times 10^{8}$ and $\sim 280 \mu K$ respectively. For magnetic trapping of atoms, the atom cloud from U-MOT needs to be further cooled and to be placed near the atom-chip surface. This can be achieved by performing G-MOT stage after U-MOT is loaded fully. To achieve G-MOT after U-MOT, the current in U-wire is ramped (in 100 ms) from 60 A to 80 A and current in Y-bias coil is ramped (in 100 ms) from from 15 A to 40 A. The cooling laser frequency is further detuned from -14 MHz  to -22 MHz. In G-MOT, atom cloud is moved vertically in z-direction to a position $\sim$ 2 mm from the chip surface as shown in Fig. 2(b). The number and temperature of cold atoms in G-MOT are $\sim 7 \times 10^{7}$ and $\sim 510\mu K$ respectively.

\begin{figure}
\centering
\includegraphics[width=8.0 cm]{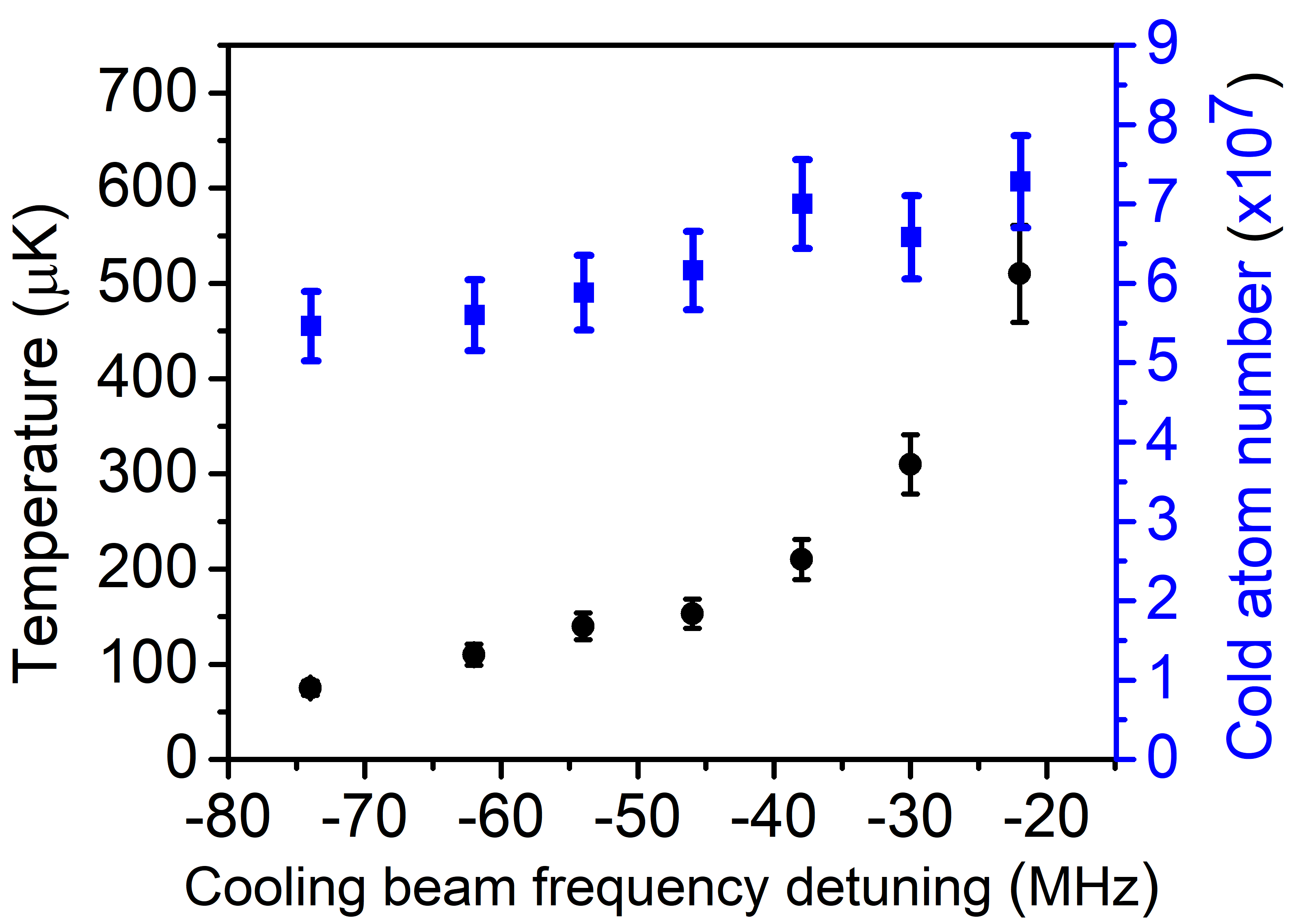}
\caption{ The variation in number (squares) and temperature (circles) of cold atoms in the G-MOT with frequency detuning of the cooling laser beam.}
\end{figure}

The temperature of the G-MOT cloud was further reduced by varying the cooling laser frequency from - 22 MHz to - 74 MHz in time duration of 10 ms. The cooling laser beam power was observed to decrease linearly with detuning at the rate $\sim$ 0.25 mW/MHz due to spectral response of AOM. As a result, the temperature of the cold atom cloud reduced from $\sim 510 \mu K$ to $\sim 77 \mu K$. However, the number of cold atoms ($\sim 5.5 \times 10^{7}$) in G-MOT remains nearly unchanged as shown in Fig. 3.\\

\begin{figure}
\centering
\includegraphics[width=9.0 cm]{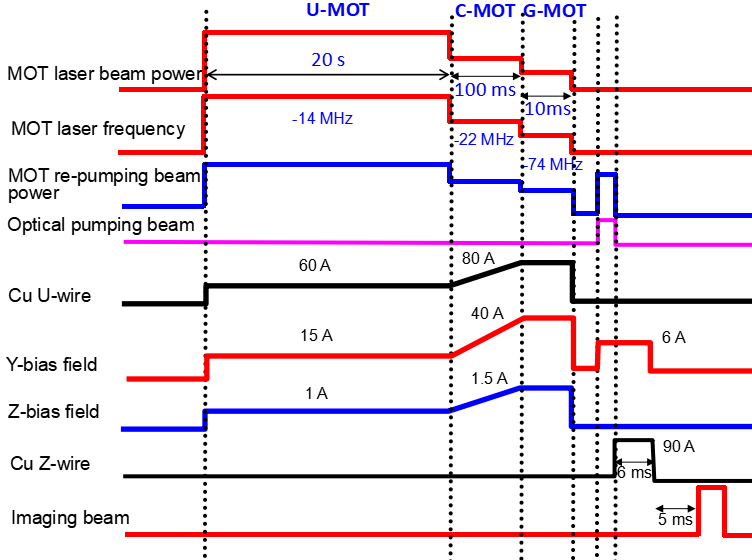}
\caption{ Schematic of the sequence of various events from U-MOT formation to optical pumping and imaging of atom cloud in Stern-Gerlach (SG) experiments. }
\end{figure}

Fig. 4 shows the experimental time sequence used for performing optical pumping experiments in presence of SG magnetic field. At the end of G-MOT stage, all the magnetic fields and laser fields are switched-off for the duration 1.5 ms. Then optical pumping beam, Y-bias field, current in Copper Z-wire (i.e. SG field gradient) and imaging beam are switched-on in a particular sequence.  The current in Copper Z-wire (90 A) is switched-on after the optical pumping of atoms, before imaging of atom cloud using fluorescence imaging technique.

\section{Stern-Gerlach Force}
In presence of a magnetic field gradient, the atoms in different Zeeman hyperfine state ($\ket{F, \, m_{F}}$) experience a force which is dependent on $m_{F}$. This force is called Stern-Gerlach (SG) force and is given as \cite{kafer} ,

\begin{equation}
F_{z}(m_{F})=\mu_{B} \, g_{F} \, m_{F} \, \dfrac{dB}{dz}  
\end{equation}

where $\mu_{B}$ is Bohr magneton, $g_{F}$ is lande's g factor, $m_{F}$ is magnetic  hyperfine angular momentum and dB/dz is magnetic field gradient (i.e. SG field gradient) along z-direction .\\

Owing to the above equation, because of state dependent acceleration, the atoms in different $m_{F}$ states in an atom cloud are expected to separate out according to their $m_{F}$ values, when they evolve in a SG magnetic field gradient \cite{stamper, marechal}. Thus the SG effect can be used to spatially separate (and detect) the population of atoms in different $m_{F}$ states. This effect has been utilized in our optical pumping experiments to identify the population in different $m_{F}$ states in an atom cloud of $^{87}Rb$ atoms cooled in F=2 hyperfine state. By accurately measuring population in different $m_{F}$ levels, the optical pumping of atoms was maximized to $m_{F}$=2 state. 

\begin{figure}
\centering
\subfigure[]{
\resizebox*{3 cm}{!}{\includegraphics{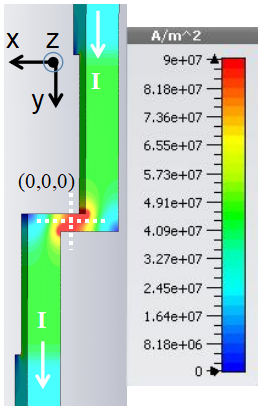}}}\hspace{0pt}
\subfigure[]{
\resizebox*{5 cm}{!}{\includegraphics{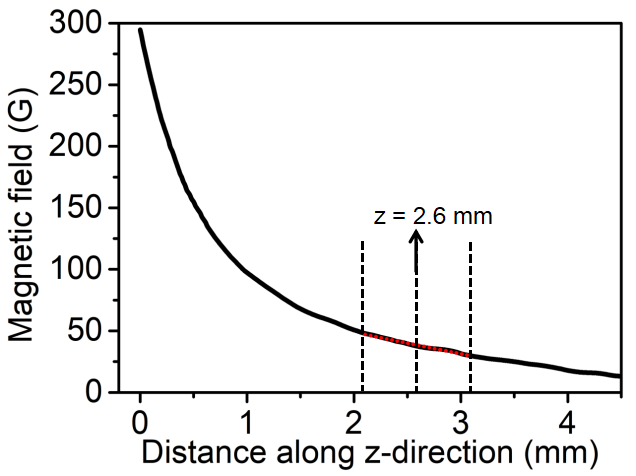}}}
\caption{ (a) The simulated current density distribution in copper Z-wire is shown with color code. (b) The variation of magnetic field due to flow of current in copper Z-wire along z-direction (0,0,z). The Z-wire dimensions at Z-joint are : 1 mm width along y-direction, 1 mm thickness along z-direction, and 4.0 mm length along x-direction). The coordinates of mid point of Z-joint are (0,0,0). }
\label{sample-figure}
\end{figure}

In our experiments, the state dependent SG force on cold atoms from G-MOT  was applied by flowing a dc current in a copper Z-wire placed behind the atom-chip. The current density variation in copper Z-wire due to flow of 90 A current is shown in Fig. 5(a). Using these current density values, the variation in magnetic field along z-direction, due to the Z-wire (carrying current of 90 A), is shown in Fig. 5(b). It can be noted here from the figure that field gradient at the G-MOT cloud position (which is $\sim$ 2.6 mm below the Copper Z-wire) is $(17.61 \pm 0.19)$ G/mm. \\

\section{Results and discussion}
We have investigated the effect of two experimental parameters, namely,  SG magnetic field pulse duration and temperature of cold atom cloud, on the $m_{F}$ state dependent splitting of cold atom cloud. Both these parameters are important in SG experiment to separate and detect group of atoms with different $m_{F}$ states (i.e. Zeeman magnetic states) in the atom cloud. These results are discussed as follows.

\subsection{Effect of SG field pulse duration}

\begin{figure}[ht]
\centering
\includegraphics[width=9.0 cm]{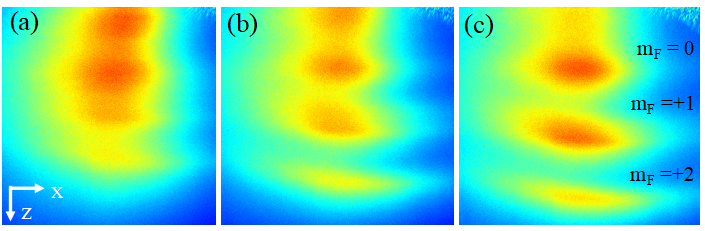}
\caption{ CCD image of the cold $^{87}Rb$ atom cloud after different values of duration $\Delta t_{SG}$ of SG field :(a) 4 ms, (b) 6 ms, and (c) 8 ms.}
\end{figure}

 The pulse duration of SG field was varied in the experiment and the atom cloud was imaged after a time of flight of 5 ms after evolution in the magnetic field gradient. Fig. 6 shows the CCD images of the cold $^{87}Rb$ atoms for different values of SG magnetic field pulse duration used in the experiment. It is evident from the Fig. 6 that as pulse duration was increased from 4 ms to 8 ms, the relative separation between the group of cold atoms in different $m_{F}$ state was also increased. For small pulse duration (i.e. 4 ms), the separation between atoms in different $m_{F}$ states was not good enough to resolve the atom clouds clearly (Fig. 6(a). With increase in SG field pulse duration (from 4 ms to 6 ms and 8 ms), the separation between atom clouds in different $m_{F}$ states was increased and clouds in different states were clearly resolved (Fig. 6 (b) and (c)). The cold atoms in $m_{F}$ = +2 state experience maximum SG force along z-axis. Therefore, this group of atoms is found at the farthest position along z-direction in the CCD image. The cold atoms in $m_{F}$ = 0 state does not experience any SG force. Therefore, this group of atoms is not affected by SG field and its position remains nearly same in all the images ((a), (b) and (c)). Since atom clouds with negative $m_{F}$ values would move in negative direction of z-axis, these clouds may be lost after colliding with atom-chip surface. That is why we do not observe any cloud for state $m_{F}$ = -2 in the images in Fig.6. \\
\begin{figure}
\centering
\includegraphics[width=6.0 cm]{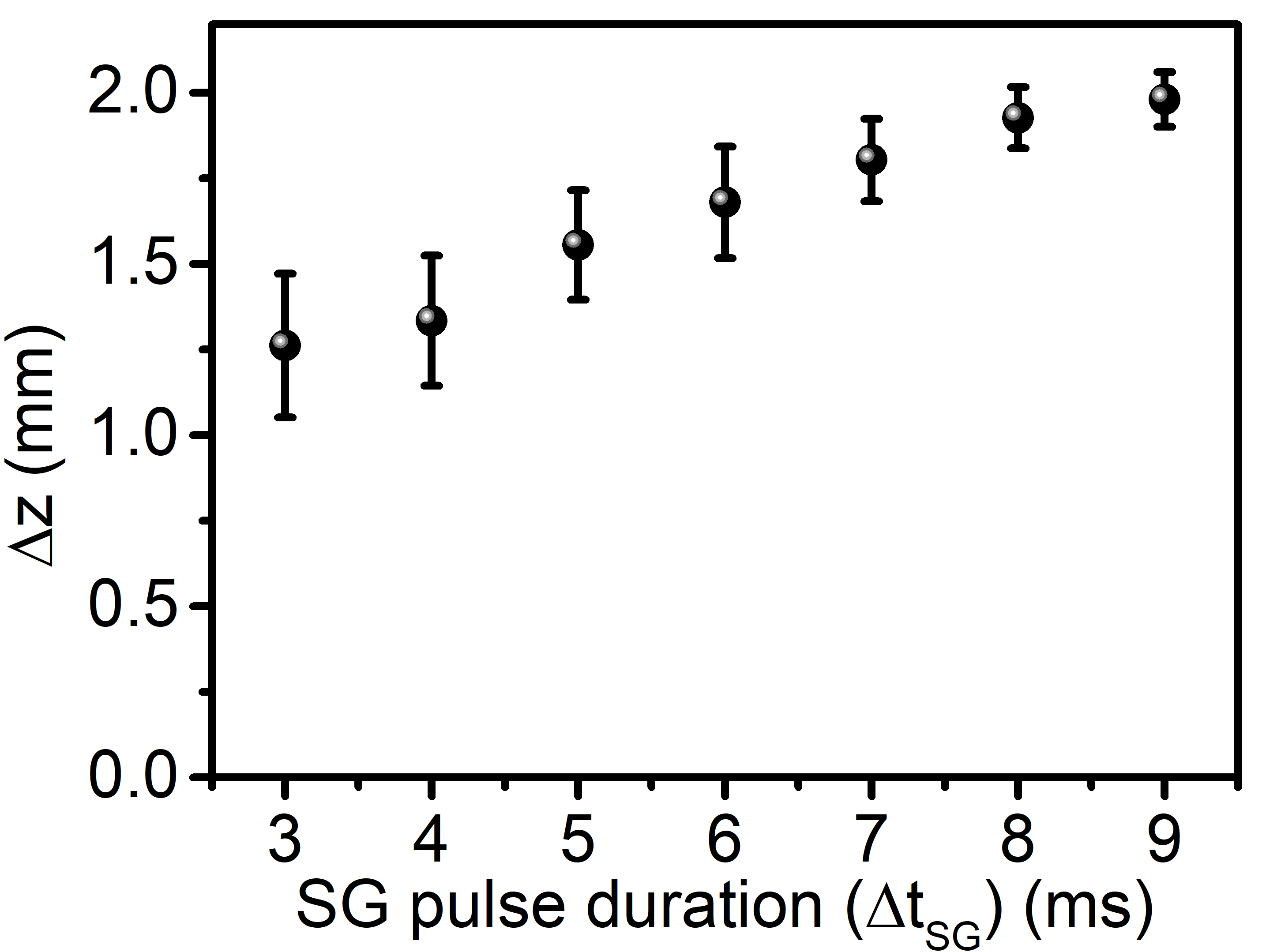}
\caption{ The measured variation in peak to peak separation ($\Delta z$) between the cold atoms in $m_{F} = + 1$ and in $m_{F} = + 2$ states as a function of SG field pulse duration for a fixed value (90 A) of current in Copper Z-wire. }
\end{figure}
Fig. 7 shows the experimentally measured separation ($\Delta z$) between the cold atoms in $m_{F} = + 1$ and in $m_{F} = + 2$ states as a function of SG field pulse duration ($\Delta t_{SG}$), at a fixed value of magnetic field gradient ($\sim$ 17.6 G/mm) due to $\sim$ 90 A current in copper Z-wire. The separation between cold atom clouds in two states ($\Delta z$) increases with ($\Delta t_{SG}$). This relative separation ($\Delta z$) increases from $\sim$ 1.2 mm to $\sim$ 2.0 mm as SG pulse duration increases from 3 ms to 9 ms respectively as shown in Fig. 7. It was observed that for $\Delta t_{SG}$ = 6 ms, the separation of atoms in different $m_{F}$ states was clearly resolved. Therefore, this pulse duration was used in the subsequent experiments.
\subsection{Effect of temperature of atom cloud}
In SG experiments, we have also studied the effect of temperature of atom cloud in G-MOT on spatial splitting of atom cloud in different parts corresponding to different $m_{F}$ states. The temperature in G-MOT was changed by changing the cooling laser beam detuning. The SG pulse duration of 6 ms and a current of 90 A in copper Z-wire was used to apply the SG force the atom cloud. The images of atom cloud were taken after 5 ms duration in time of flight after the SG field pulse. Fig. 8 shows the CCD image of the $^{87}Rb$ atom cloud after splitting in different $m_{F}$ states for different initial temperature of cloud in G-MOT. 

\begin{figure}[ht]
\centering
\includegraphics[width=8.0 cm]{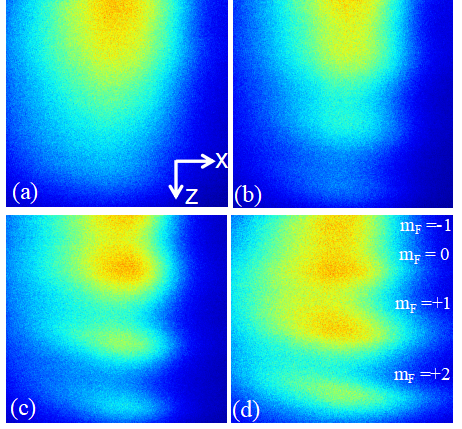}
\caption{ The images of split atom cloud of $^{87}Rb$ atoms after SG splitting. The temperature of cloud in G-MOT was kept at different values for these images : (a) $\sim$ for 510 $\mu K$, (b) $\sim$ 210 $\mu K$, (c) $\sim$ 130 $\mu K$, (d) $\sim$ 77 $\mu K$. The SG pulse duration was 6 ms at 90 A current in copper Z-wire, and atom cloud images were taken after 5 ms of time of flight after the SG field pulse.}
\end{figure}

It is evident from images in Fig.8 that lower temperature of initial atom cloud improves the spatial resolution of cloud split in different states. The higher initial temperature of cloud from G-MOT results in faster expansion of cloud in SG field, which keeps the split parts overlapped as shown in Fig. 8(a). The group of atoms in different $m_{F}$ states after SG splitting are better resolved at lower initial temperature of cloud (e.g. Fig. 8 (c) and (d)). \\

\subsection{Optimization of optical pumping process}

The SG state detection technique has been finally utilized to maximize the optical pumping efficiency for preparing the $^{87}Rb$ atom cloud in the trappable Zeeman hyperfine state ($\ket{ F=2, \, m_{F} = + 2}$ ). All magnetic fields and laser beams were switched-off after G-MOT stage for a duration of 1.5 ms. A right handed circularly polarized ($\sigma ^{+}$) optical pumping beam pulse resonant to $D_{2}$-line transition $F=2 \xrightarrow{} F'=2$ was directed along y-direction to pump atom cloud to Zeeman hyperfine state ($ \ket{F=2, m_{F} = + 2}$ ) in presence of an axial magnetic field of $\sim$ 3.6 G.  A re-pumper laser beam pulse resonant to $D_{2}$-line transition $F=1 \xrightarrow{} F'=2$ was also overlapped with the optical pumping beam to avoid accumulation of atoms in F=1 state. Atoms are expected to be in ($\ket{ F=2, \, m_{F} = + 2}$ ) state after application of optical pumping pulse. After optical pumping, SG state detection was performed by applying a current pulse of 90 A (6 ms) to copper Z-wire to generate Stern-Gerlach (SG) force on atoms. Then after 5 ms of time of flight, the cloud was imaged by fluorescence technique to detect the population of atoms in different Zeeman states.\\

\begin{figure}[ht]
\centering
\includegraphics[width=9.0 cm]{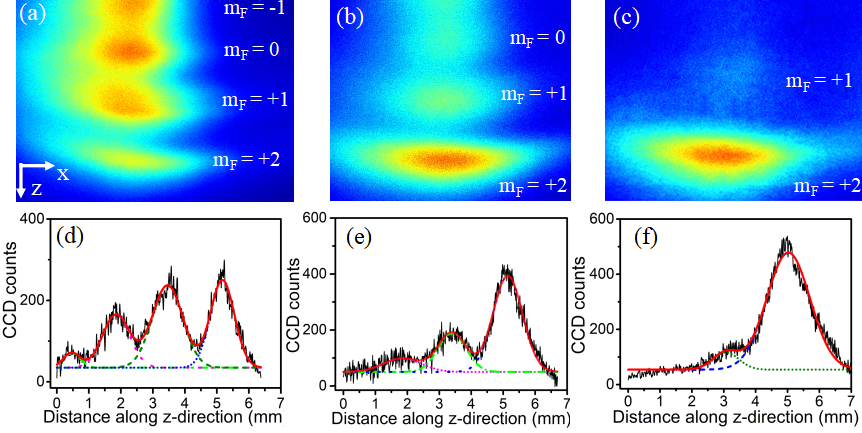}
\caption{ The CCD images of atom cloud under Stern-Gerlach (SG) force showing the distribution of population of $^{87}Rb$ atoms in different $m_{F}$ state: (a) without optical pumping, (b) with optical pumping pulse of duration $\sim$ 300 $\mu s$ , and (c) with optical pumping pulse of duration $\sim$ 500 $\mu s$. The atom clouds in different $m_{F}$ states ($m_{F} = -1, 0, +1, +2$) are shown in the images. The plots (d)-(f) show the line profiles along z-direction for images (a)-(c) respectively.}
\end{figure}


 Fig. 9 shows the spatially separated cold atom clouds in different $m_{F}$ states and the corresponding line profiles along z-direction. In the absence of optical pumping beam (Fig.9 (a) and (d)), peaks correspond to different states ($\ket{F=2, m_{F} = -1, 0, +1, +2}$). The cold atoms in $(F=2, m_{F} = -2)$ states are not seen in the image as they are accelerated upwards and lost to the atom-chip surface. We noted that without optical pumping, in split cloud after SG pulse, $\sim$ 24 $\%$ of atoms were present in $\ket{F=2, m_{F} = + 2}$ state, with respect to number of atoms in G-MOT. With use of optical pumping, the population in  $\ket{F=2, m_{F} = + 2}$ state was increased as shown in Fig.9 (b) and (c). After using an optimized optical pumping pulse ($\sim$ 500 $\mu s$,  power = 300 $\mu W$) in presence of $\sim$ 3.6 G bias magnetic field, $\sim$ 92 $\%$ of G-MOT atoms were transferred to $\ket{F=2, m_{F} = + 2}$ state. Any further change in power or pulse duration of optical pumping beam did not help in improving optical pumping efficiency significantly.\\ 
 
\subsection{Magnetic trapping of atoms}

\begin{figure}[htp]
\centering
\includegraphics[width= 8.5 cm]{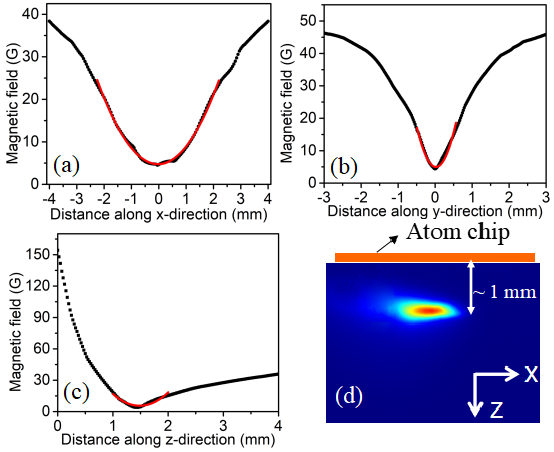}
\caption{ The spatial variation in magnitude of the net magnetic field due to copper Z-wire and bias coils. The current in the copper Z-wire was $I_{Z} = 60 A$ and magnetic fields due to bias coils were $B_{y}$ = 38 G and $B_{x}$ = -25 G. The image in (d) is the fluorescence image of magnetically trapped atom cloud.}
\end{figure}

After optical pumping of atoms ( with pumping pulse of optimized duration of  $\sim 500 \mu s$ ) to Zeeman hyperfine quantum state ($(\ket{F=2, \, m_{F} = + 2})$), the magnetic trapping of atoms was performed. For this, all lasers and magnetic fields were switched-off after optical pumping. Then trap fields were switching-on within 1 ms time. Nearly 60 A of dc current was switched-on in the copper Z-wire ($I_{Z} = 60 A$) along with the bias magnetic fields of strength $B_{y}$ = 38 G and $B_{x}$ = -25 G. Fig. 10 shows the calculated magnetic field strength due to Z-wire and bias fields for the above parameters.  Fig. 10 (a), (b) and (c) show the calculated absolute magnetic field (or potential) profiles along x-, y- and z-direction respectively. The trap configuration is close to Ioffe-Pritchard configuration \cite{esslinger}. The trap frequencies at this current are $\omega_{x}/2\pi \sim 36$ Hz and $\omega_{y,z}/2\pi \sim 110$ Hz. Due to asymmetry in the trap frequencies, the trapped atom cloud is expected to be elongated along x-direction, as shown in Fig. 10(d). The CCD image in Fig. 10 (d) shows the fluorescence image of magnetically trapped atom cloud in X-Z plane. The total number of cold $^{87}Rb$ atoms in the magnetic trap was $\sim 1 \times 10^{7}$ out of $\sim 5 \times 10^{7}$ in G-MOT, leading to trapping efficiency $\sim 20 \% $. We believe that this trapping efficiency can be improved by improving several factors such as switching-on magnetic trap slowly, lowering the temperature of atom cloud in G-MOT, etc. \\

\section{Conclusion}
To conclude, Stern-Gerlach effect of magnetic field on cold $^{87}Rb$ atoms has been used to spatially separate and detect atoms in different Zeeman hyperfine quantum states. It is found that the spatial separation of cold atoms in different magnetic hyperfine states depends on SG field pulse duration and temperature of atom cloud temperature. Using this SG state detection technique, the optimization of optical pumping of cold $^{87}Rb$ atoms to a trappable Zeeman hyperfine state has been demonstrated. Nearly 92 $\%$ of cold atoms from grey magneto-optical trap (G-MOT) were optically pumped to $\ket{F=2, m_{F} = + 2}$ state using this SG technique.

\begin{acknowledgments}
We are grateful to R. Folman and Omer Amit, Ben-Gurion University, Israel for technical suggestions during experiments. We are also thankful to our colleagues S. P. Ram, S. Singh and K. Bhardwaj, LPAS, RRCAT for their help during the experiments. The technical assistance provided by Amit Chaudhary, LPAS, RRCAT during the experiments is also gratefully acknowledged.
\end{acknowledgments}
\bibliography{reference-vivek}

\end{document}